\documentclass[twocolumn,showpacs,preprintnumbers,amsmath,amssymb]{revtex4}
\usepackage{graphicx}
\usepackage{dcolumn}
\usepackage{bm}
\DeclareGraphicsExtensions{.pdf,.png,.jpg}

\begin{document}
 
\title{Origin of structural and magnetic transitions in BaFe$_{2-x}$Ru$_x$As$_2$ materials}
\author{Smritijit Sen, Haranath Ghosh\footnote{Corresponding author : hng@rrcat.gov.in}, A. K. Sinha and A. Bharathi $^1$.}
\affiliation{ Indus Synchroton Utilization Division, Raja Ramanna Center for Advanced Technology, 
Indore -452013, India. \\
$^1$ Materials Science Group, Indira Gandhi Centre for Atomic Research, Kalpakam, India. 
}
\date{\today}
\begin{abstract}
 Using the experimentally measured temperature and doping dependent structural
 parameters on Ru doped BaFe$_2$As$_2$, orbital-dependent reconstruction of        
 the electronic structure across the magnetostructural transition is found, through first
 principle simulations. Below structural transition there exists two distinct Fe - Fe bond 
 distances which modifies the Fe-d$_{xy}$ orbital largely due to its planar spatial
 extension leading to Lifshitz transition, while the otherwise degenerate Fe-d$_{xz}$
 and d$_{yz}$ orbitals become non-degenerate, giving rise to orbital order. The orbital order
 follows the temperature dependence of orthorhombocity and is also the cause of two distinct
 Fe - Fe bond distances. Doping dependent Fermi surfaces show nearly equal expansion of both the
 electron and hole like Fermi surfaces whereas the hole Fermi surface shrinks with
 temperature but the electron Fermi surface expands comparatively slowly. The
 observed structural transition in this compound is electronic in origin, occurs
 close to the Lifshitz transition whereas the suppression of the concurrent magnetic
 transition is due to loss of temperature dependent nesting of Fermi surface.

\vspace{1pc}
\end{abstract}
\maketitle


Discovery of superconductivity in a plethora of Fe based compounds has been 
significant to the history of superconductivity, as it bears strong 
similarity with oxide superconductors, in terms of phase diagrams, but their 
properties are  fundamentally different \cite{Kamihara}.
These differences include its superiority in technological applicability due to high critical
current at high fields \cite{application} over the other high temperature superconductors;
apart from various unconventional properties \cite{stewart} like different
fermiology, BCS characteristic ratio, jump in specific heat proportional to T$_c{^3}$, no
oxygen isotope shift (but Fe), linear temperature dependence of spin susceptibility, scaling of
spin resonance with T$_c$, structural and magnetic transition etc. While the mechanism of
superconductivity in Fe-based materials is still unknown, a few things mentioned above are
consistently observed. The whole family of Fe-based
materials may be broadly classified into six categories e.g, 
1111 (like LaOFeAs), 122 (like BaFe$_2$As$_2$), 111 (like LiFeAs),
11 (like FeSe), 122* (A$_x$Fe$_{2-y}$Se$_2$, A = K, Rb, Cs), 21311 (Sr$_2$ScO$_3$FeP); 
among them 122,11 materials structural and 
magnetic transitions occurring at the same temperature,
 whereas in 1111 and 122*
they occur at different temperatures \cite{Luetkens,Huang,Li,Parkar,Liu}. 
Proximity of superconducting phase
to magnetic and structural transitions
indicates possible influence on the former due to the later
\cite{Cruz,Pratt,Drew,Mazin,Dong,orbital}. The magnetic spin
density wave (SDW) state appear due to antiferromagnetic like arrangement of Fe spins and
nesting of Fermi surface; there are growing evidences that the structural transition has also an
electronic origin \cite{nematic}.

 \begin{figure}
\flushleft\includegraphics[height=8.0cm,width=8.0cm]{fig1-prl.eps}
\includegraphics[viewport=345 200 120 20,scale=0.650]{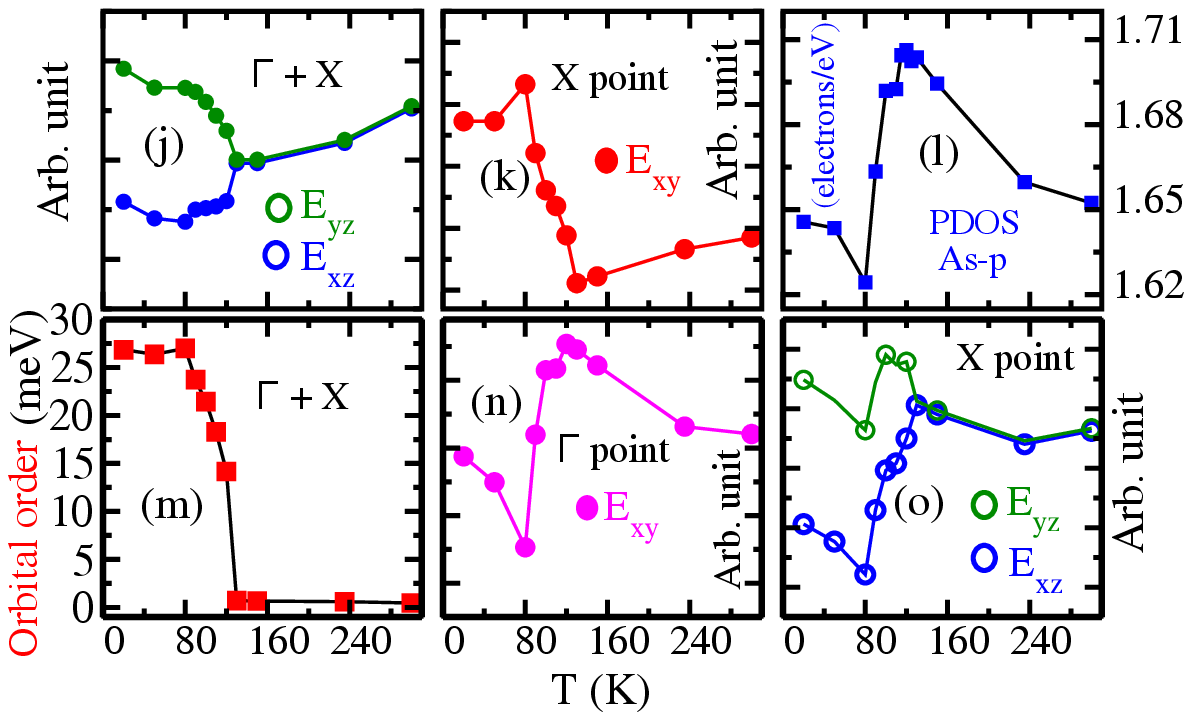}
\vskip 1.6in
 \caption{Experimental temperature variation of {\bf (a)} z$_{As}$ {\bf (b)} Fe-As bond distance
 {\bf (c)} orthorhombicity $\delta$ of BaFe$_{1.9}$Ru$_{0.1}$As$_2$ from ref \cite{arxiv}
 The orthorhombic distortion occurs rapidly after the high temperature tetragonal
 phase transforms into low temperature orthorhombic phase around 125 K. Theoretically simulated
 thermal variations of {\bf (d)} total density of state at the Fermi level,
 {\bf (e)} difference of up and down electron density of states at the Fermi level
  and {\bf (f)} magnetic moment in the second column. Thermal variations of density of
states of Fe {\bf (g)} up {\bf (h)} down electrons. {\bf (i)}
The total converged energy as a function of temperature also follows to that of the $z_{As}$,
Fe-As and justify the behavior of DOS. {\bf (j)}, Sum of the energies of the d$_{xz}$ at $\Gamma$
and X point and the same for d$_{yz}$ as a function of temperature is the cause of two Fe-Fe-bond 
distances. {\bf (m)} The orbital order as a function of temperature (see text for details). 
{\bf (k)} Energies of the d$_{xy}$ band as a function of temperature at X and {\bf (n)} $\Gamma$ points.
{\bf (l)} Partial density of states of As as a function of temperature. {\bf (o)} Energies of
d$_{xz}$ and d$_{yz}$ of Fe as a function of temperature at the X-point}

\label{figstructure}
\end{figure}

\begin{figure}
\centering
\includegraphics[viewport=0890 650 120 20,scale=0.23]{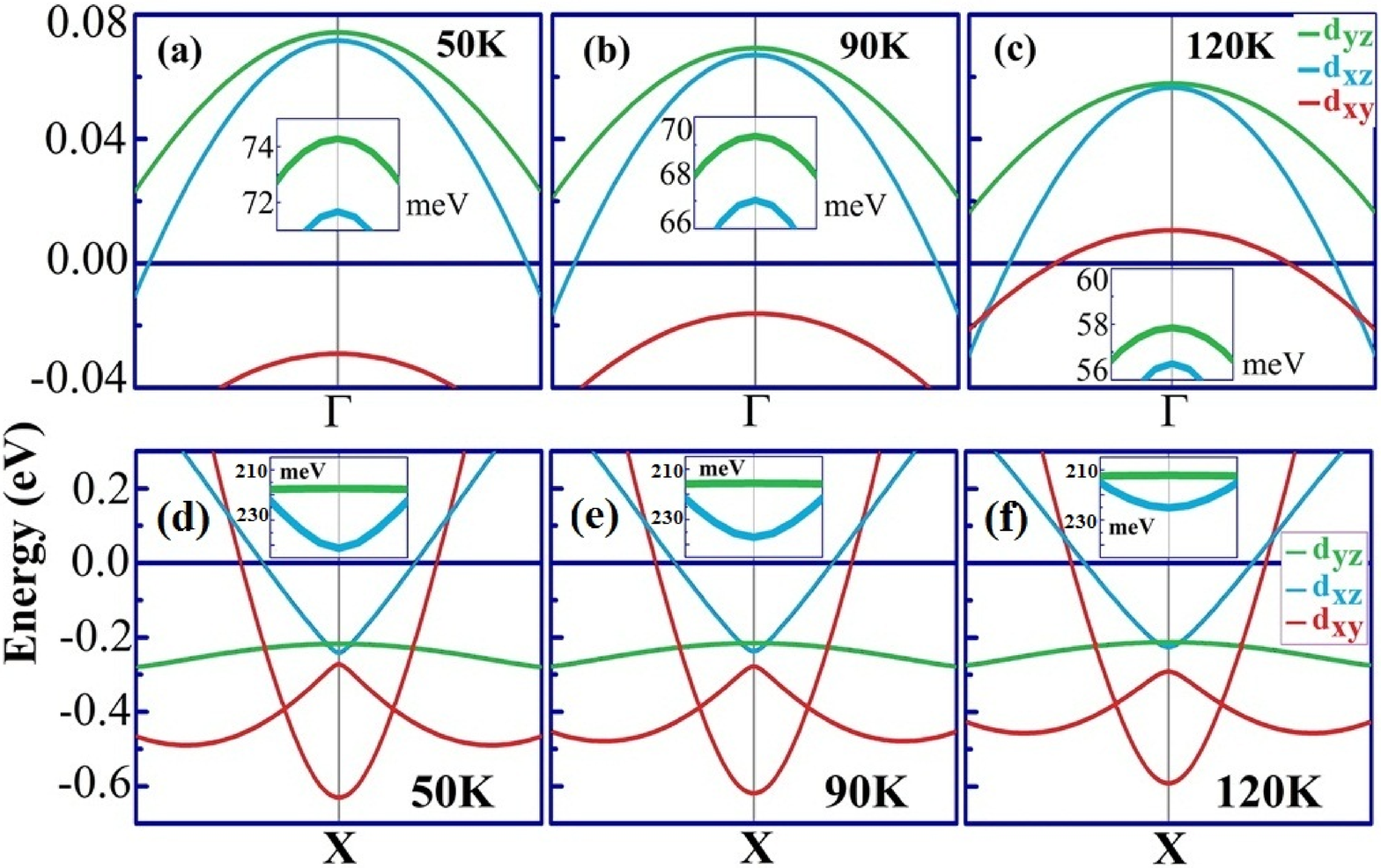}
\includegraphics[viewport=0130 1300 120 20,scale=0.23]{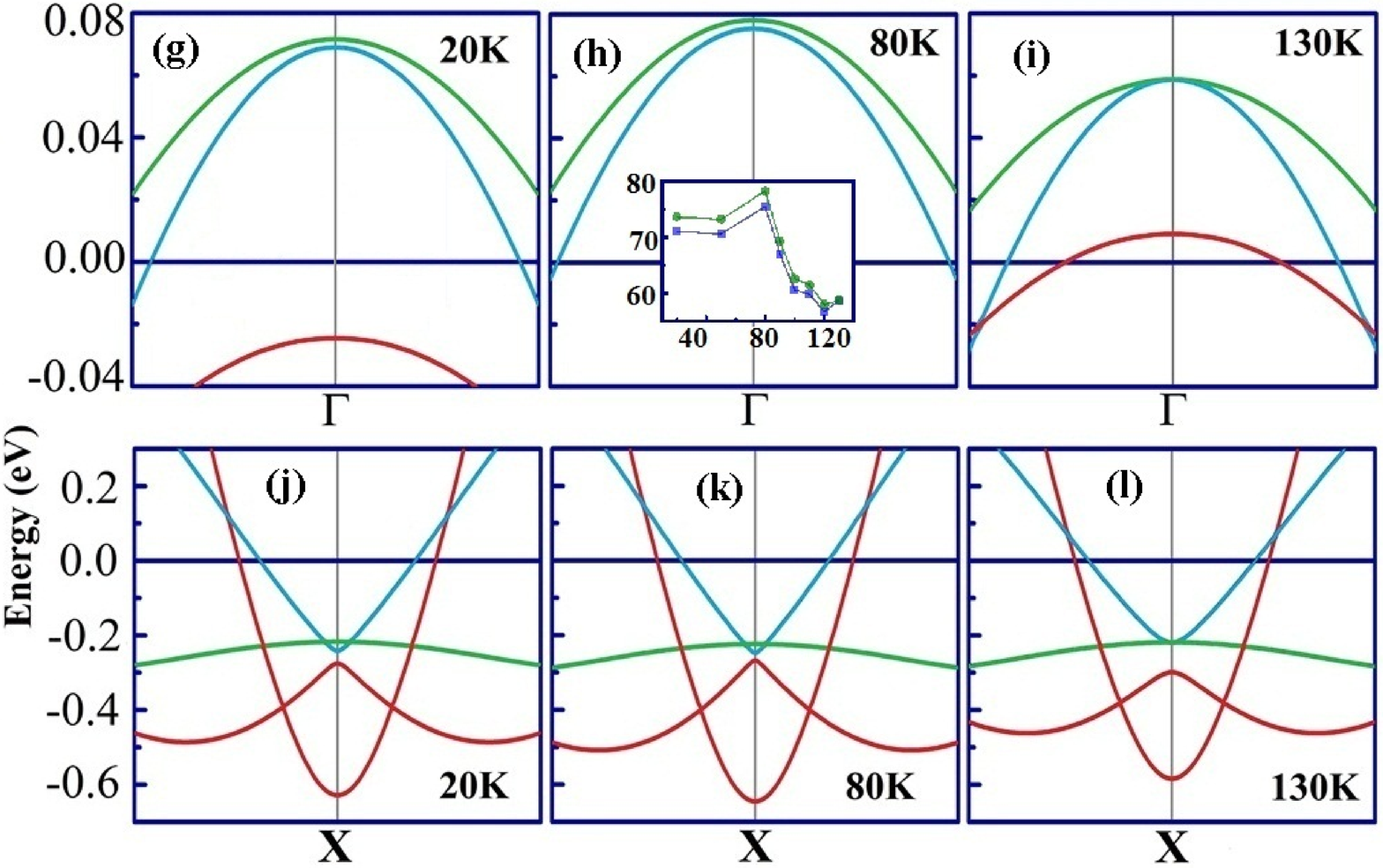}

\vskip 4.1in

\caption{Band structure around $\Gamma$ (first and third row) and 
X-point (second and forth row) respectively at
different temperatures. Lifshitz transition is observed as the d$_{xy}$ orbital moves down 
the Fermi level [see figures {\bf (a),(b),(g), (h)}]. Splits between the d$_{xz}$ and d$_{yz}$ orbitals
are presented by insets in the figures ({\bf a, b, c, d, e, f}), 
corresponding energies at the $\Gamma$ point is shown in the inset {\bf (h)} (see also the same at 
X point in the lowest row of Fig. \ref{figstructure}). 
}
\label{band-temp}
\end{figure}
Structurally, building block of all families of Fe-based materials is Fe-pnictogen/chalcogen
layers in which the pnictogens/chalcogens are slightly above or below the Fe-plane.
 There exists a definite correlation among
structural lattice parameters like $z_{As}$ (anion height), bond lengths (Fe-Fe and Fe-As),
As-Fe-As bond angles and superconducting transition temperature $T_c$ in all families of
Fe-based superconductors \cite{Mizuguchi,Lee,Zhao}.
In particular, superconducting transition temperature (T$_c$) is very closely related to anion 
height from Fe layer (directly related to $z_{As}$). For most of the Fe-based families like 
1111, 122, 111 and 11 the anion height as a function of T$_c$ follows a universal trend in 
ambient pressure as well as under high pressure \cite{Mizuguchi}. Relation among
 Fe-Fe and Fe-As bond distances with T$_c$ is also available in the 
literature \cite{Zhao}.
 The bond angle of As-Fe-As is also related to T$_c$ as the distortion of
FeAs$_4$ tetrahedron reduces T$_c$ and maximum T$_c$ is obtained when the FeAs$_4$ tetrahedron
is perfectly regular \cite{Lee,Shirage}. All these structural
 parameters described above 
are very sensitive functions of temperature, doping etc. Any microscopic understanding on
origin of various
temperature dependent experimental observations ({\it e.g}, temperature dependent angle resolved 
photo emission (ARPES) studies etc.
\cite{DhakaT}) calls for temperature dependent first principle studies. However, first principle
studies which evolve from solution of many body Schr\"{o}dinger equation cannot account for such
temperature dependencies. On the other hand, density functional theory has failed
to produce optimized structures reproducing experimental values of $z_{As}$ (even if one uses
temperature dependent basic lattice parameters $a$(T), $b$(T), $c$(T)), a parameter which
is found to be extremely crucial in reproducing other experimentally observed 
structural parameters and
associated physical properties \cite{DJSingh,arxiv,dft,zAs}. 
Therefore, hybridization of experimental inputs of temperature/doping/pressure dependent 
basic lattice parameters along with $z_{As}$ in density functional theory would be a 
state-of-the-art first principle approach for understanding experimental observations on Fe-based 
materials.
For example, the temperature dependent ARPES data by Dhaka {\it et al}., \cite{DhakaT} could not
be analyzed by themselves satisfactorily because of absence of temperature dependent 
crystallographic structural parameters (like z$_{As}$).

In this letter, using temperature and doping dependent lattice parameters 
$a$($x$,T), $b$($x$,T), $c$($x$,T) and z$_{As}$($x$,T) obtained from Synchrotron 
radiation
X-ray diffraction studies on Ru doped BaFe$_2$As$_2$ as inputs,
we show that the results of first principle simulations reproduce experimentally observed ARPES
data which so far remained unexplained.
   In particular, we demonstrate through first principle simulations,
 that the temperature dependent pnictide height (z$_{As}$(T))
plays a very crucial role in structural and magnetic transition.
Temperature dependencies of the electronic structure closely follow
that of the z$_{As}$(T) for $x$= 0.1 and is essential in explaining the 
temperature dependent band shifts observed in ARPES studies. 
We further show, both experimentally and  theoretically,
that below the structural transition there
exists two distinct Fe - Fe bond distances which modify the 
Fe-d$_{xy}$ orbital largely due to its planar spatial extension leading 
to Lifshitz transition \cite{lifshitz}, whereas the Fe-d$_{xz}$ and 
d$_{yz}$ orbitals become 
non-degenerate, which were degenerate above the structural transition 
giving rise to orbital order.
We establish that the orbital order follows the temperature dependence of 
the experimentally determined orthorhombocity indicating electronic nature 
of the structural transition (see Fig. \ref{figstructure}).
On the other hand, the two distinct Fe - Fe bond distances (which would
correspond to two different exchange couplings \cite{yili,magno}), is 
 a consequence of complex orbital order and also 
marks the appearance of the magnetic ground state as evidenced through the 
temperature dependence of the net difference
in the up and down spin electronic density of states at the Fermi level \---
resulting in a simultaneous electronic magneto-structural transition.
Doping dependent Fermi surfaces show nearly equal expansion of both 
the electron and hole like Fermi surfaces upto 40 $\%$ doping
whereas the hole Fermi surface shrinks with temperature but the electron 
Fermi surface expands 
comparatively slowly. Therefore, the observed structural transition in this
compound is electronic in origin, occurs close to the Liftshitz transition 
whereas the suppression of the concurrent magnetic transition is due to 
loss of temperature
dependent nesting of Fermi surface. Futhermore,
we show that by employing temperature and doping dependent basic
lattice parameters $a$(T,$x$), $b$(T,$x$), $c$(T,$x$) and the $z_{As}$ (T,$x$) 
a satisfactory explanation to the
observed data by Dhaka {\it et al}., can be obtained. Claims made in this paragraph  
above are demonstrated in the figures 1 -4 of the rest of the letter.

\begin{figure}
\includegraphics[scale=0.30]{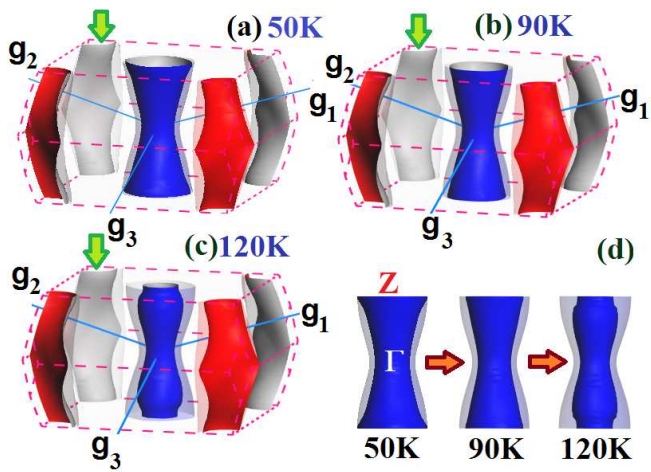}
\caption{Theoretically calculated three dimensional Fermi surfaces at different temperatures. 
Shrinking of hole like FS towards higher temperatures as well as surging of electron like FS leading
to loss of nesting of FS is appreciable. 
}
\label{fs-temp}
\end{figure}

\begin{figure}
\flushleft\includegraphics[width=0.47\textwidth,height=0.40\textwidth]{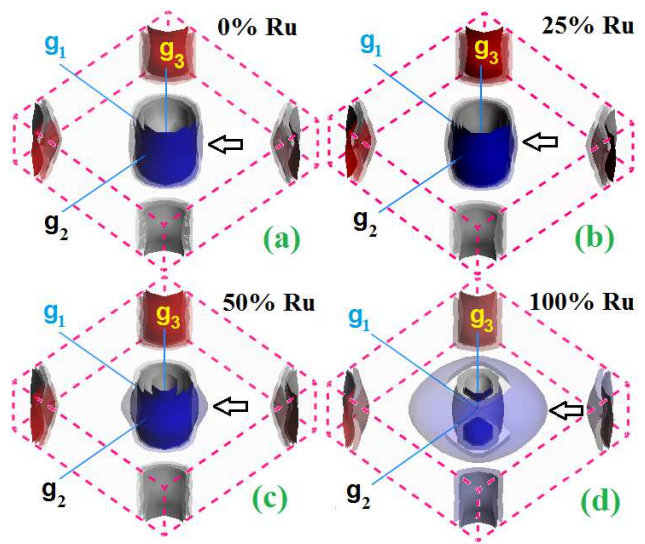}
\caption{Theoretically calculated three dimensional Fermi surfaces at different Ru doping 
concentrations (as indicated in the figure). Upto about 40 $\%$ doping the nesting of the FS
remains preserved. For BaRu$_2$As$_2$ compounds the simulated FS do not show nesting. 
}
\label{fs-dope}
\end{figure}

Temperature dependent Rietveld quality data specially on $z_{As}$ of any of the Ba122 system is rare.
 Recently, such high quality Rietveld data established relationship among
various structural lattice parameters like $z_{As}$, bond lengths (Fe-Fe and Fe-As), to the structural
transition temperature in 5$\%$ Ru doped BaFe$_2$As$_2$ \cite{arxiv} material.
Using temperature dependent and doping dependent 
experimental lattice parameters $a$ (T,$x$), $b$ (T,$x$),
$c$ (T,$x$) and $z_{As}$(T,$x$) \cite{arxiv} as inputs in our first principle
simulations
we obtain electronic band structure, density of states, Fermi surfaces as a 
function of temperature as well as
doping, to explain the observed anomalies microscopically.
Our first principle {\it ab-initio} simulations of electronic structure calculations 
are performed employing Material Studio 7.0, CASTEP package \cite{CASTEP} which exploits 
the plane-wave pseudopotential method based on density functional theory (DFT).
In all of our calculations the electronic exchange correlation
is treated within the generalized gradient approximation (GGA) using Perdew-Burke-Enzerhof 
(PBE) functional \cite{PBE}. Tackling small fraction of Ru substitution in place of Fe 
is accomplished by considering virtual crystal approximation (VCA) based on the Mixture 
Atom Editor of CASTEP
program in Material Studio 7.0,
as well as super-cell approach.
 Spin polarized constrained optimization and single point energy calculations are 
performed using anti-ferromagnetic spin-stripe
 configuration \cite{AFM3} for the low temperature orthorhombic phase with space group
 symmetry Fmmm (No.69) using ultrasoft pseudopotentials and plane wave basis set with
 energy cut off 400 eV and self consistent field (SCF) tolerance $10^{-6}$ eV/atom. Brillouin zone is
 sampled in the k space within Monkhorst-Pack scheme and grid size for SCF calculation is $16\times
 16\times7$. Non-spin polarized and spin polarized calculations are performed for high temperature
 tetragonal phase with space group symmetry I4/mmm (No.139) using ultrasoft pseudopotentials and plane
 wave basis set with same energy cut off and SCF tolerance as above. Brillouin zone is
 sampled in the k space within Monkhorst-Pack scheme and grid size for SCF calculation is $16\times
  16\times5$.
 
In Fig. \ref{figstructure} there are three columns, in the first column experimentally
observed structural parameters are presented (except the fourth and fifth rows).
 In the second and third columns
theoretically computed results are presented.
While it is clear from Fig \ref{figstructure} (a) and (b) that both the $z_{As}$(T) 
and Fe-As (T) follow nearly the same
temperature variation; structural transition is marked by  the evolution 
of the orthorhombocity parameter
$\delta$, which occurs exactly at the same temperature where both
$z_{As}$(T) and Fe-As (T) show an anomalously peaked value. Values of $z_{As}$ or
Fe-As bond lengths are nearly same at very low and high temperatures but show
a very rapid temperature variations in between 80 to 125 K indicating the structural
change. Almost same temperature dependencies are seen in figures 1 (d), (g), 
(h) and (i) which represent respectively the sum total of electronic
density of states of up and down spins at the Fermi level, partial DOS of 
Fe up spins,
partial DOS of Fe down spins, and the total ground state energy of the 5 $\%$
Ru doped BaFe$_2$As$_2$ system obtained through DFT simulations. 
(The same temperature 
dependence is also followed by the As-$p$ orbitals, shown in Fig. \ref{figstructure} {\bf (l)}.) It is
also the same temperature where $z_{As}$ or Fe-As show anomaly, below which
there exists two distinct Fe-Fe distances \cite{arxiv} (see Fig. \ref{figstructure} (c), (f) and
(j)). In figure 1(e),
difference in the  DOS of up and down spins (scaled as 10$ ^{4}$)
as a function of temperature is presented. This observation correspond to AFM transition
because just above the AFM transition
$N_{\uparrow}(E_F) - N_{\downarrow}(E_F) \neq$ zero whereas it is zero inside the AFM phase
 clearly indicates the onset of AF magnetic transition. The same behavior in the temperature
dependence of the calculated net magnetic moment of the unit cell is also found (not shown 
here). In figure 1 (f) we show that there exists 
two distinct Fe-Fe bond distances exactly below the magnetic and structural 
transition --- the two distinct Fe-Fe distances are quite robust and is observable even 
in case of non-magnetic calculations. Two distinct Fe-Fe bond distances would correspond 
to two distinct exchange 
coupling constants,
a scenario observed earlier by Yildirim \cite{yili}. Furthermore, an important noticeable 
feature in all the temperature dependent structural parameters (see figures 1 (a), (b), (c), (i))
is that below T $\sim$ 80 K all the parameters
increase with lowering in temperature to reach values closer to that of the room temperature one,
the exactly same behavior is also seen in the thermal behaviors of the density of states.
Therefore, since it is the modifications in the temperature
dependent electronic density of states that correlates with all the structural lattice
parameters, the associated structural transition is electronic in origin.
We discuss further on the same below.

Figures of the lowest two rows of Fig. \ref{figstructure} are obtained from detailed temperature 
dependent electronic band structure
calculations, a glimpse of which are shown in Fig. \ref{band-temp}. 
In figure \ref{band-temp} electronic band structures around 
$\Gamma$ (1st/3rd row) and X (2nd/4th row) points are presented for different 
temperatures. One of the most important observations from the band structures
around $\Gamma$ point is that the Fe-d$_{xy}$ orbital width increases with
temperature so much (which will cause modifications in its occupation) that 
at 120 K it crosses the Fermi level. The observation that the Fe-d$_{xy}$ 
level going below the Fermi level at T= 90 K, 50 K gives rise to
 Lifshitz transition \cite{lifshitz1}.
Such temperature dependent modifications in the electronic bands crossing the Fermi
level, enhancement in the widths, are source of
 orbital fluctuations. The tip of the d$_{xy}$ band around $\Gamma$ and X-points are 
shown in Fig. \ref{figstructure} {\bf (k)}, {\bf (n)} respectively which actually 
follows the temperature dependencies of As-Fe-As angles. The tip of the d$_{xz}$,
d$_{yz}$ bands at the $\Gamma$/X point is degenerate at T = 125 K becomes
non-degenerate in the orthorhombic phase, causing an orbital ordering 
between the orbitals (d$_{xz}$, d$_{yz}$) and two Fe-Fe distances 
(see Fig. \ref{figstructure} {\bf (j)} and follow below). Such non-degenerate d$_{xz}$, d$_{yz}$ 
bands are observed experimentally recently \cite{sci-rep}. We define orbital order as, 
$ <O> = \sum_{i= \Gamma, X}$ E$_{d_{xz}}(i)$ - E$_{d_{yz}}(i)$ which are presented 
as a function of temperature
in the first two figures of the bottom row of Fig. \ref{figstructure} {\bf (m)}. 
Needless to say that the temperature dependence 
of the orbital order $<O>$ reproduces that of the experimentally determined
 orthorhombocity ($\delta$, compare figures {\bf (c)} and {\bf (m)}) 
indicating orbital ordering is the principal origin of the structural transition. 
The temperature dependence of the tip of the d$_{xz}$ 
and d$_{yz}$ (shown in Fig. \ref{band-temp}) bands at X-point given by E$_{xz}$ and 
E$_{yz}$ respectively follow the temperature dependence 
as that of the z$_{As}$ (cf. \ref{figstructure} {\bf (o)}).
Sum of the energies at X and $\Gamma$ points of d$_{xz}$ and d$_{yz}$ bands respectively are 
presented in Fig. \ref{figstructure} {\bf (j)} showing its temperature dependence similar to
 as 
that of the two distinct Fe-Fe distances in Fig. \ref{figstructure} {\bf (f)}. 
 These remarkable results clearly show that the 
structural transition dictated by $\delta$, two Fe-Fe distances, anion height z$_{As}$,
are orbital driven. Experimentally observed temperature
dependencies of z$_{As}$, Fe-As bond distance, orthorhombocity parameter $\delta$ 
are consequences of the temperature dependent modifications in the electronic structure
and vice versa. This naturally supports
nematic scenario in Fe-based superconductors \cite{nematic}.

On the other hand, the feature that the tip of the d$_{xz}$, d$_{yz}$ bands 
approach the Fermi level and that there is about 25 meV shift of
the d$_{xz} (\Gamma)$, d$_{yz} (\Gamma)$ downwards to the Fermi energy 
from 20 K to 125 K are consistent with ARPES studies of Dhaka {\it et al}.,. It should 
further be noted that such temperature dependence was not achievable when only 
thermal expansion of lattice parameters were considered in their DFT simulations.
Furthermore, widths of the d$_{xz}$, d$_{yz}$ bands around $\Gamma$ point (evaluated 
along X--$\Gamma$--X path)  decreases with temperature whereas the same around 
X-point (evaluated along $\Gamma$--X--$\Gamma$ path) increases (comparatively) slowly.
This causes the d$_{xz}$, d$_{yz}$ bands crossing the Fermi level at a shorter 
$\Delta k$ around $\Gamma$ point whereas at a somewhat larger $\Delta k$ around 
the X -point. This makes the hole Fermi surface around the $\Gamma$ point shrink 
whereas that around X-point expand a bit with temperature (see Fig. \ref{fs-temp}) 
causing temperature dependent loss of Fermi surface nesting. 
This is the reason for the decrease in hole Fermi surface
radius around the Z-point (also $\Gamma$ point) with temperature.
This naturally explains the momentum distribution curves obtained in ARPES
studies \cite{DhakaT}. Therefore, the orbital order that causes structural transition also
causes damage to the FS nesting, suppressing SDW and thus both are inter coupled \cite{ghosh}.
In Fig. \ref{fs-dope} fermiology 
of Ru doped Ba122 systems are presented, upto 40 $\%$ (which is already in tetragonal
phase) Ru substitution, nesting of the Fermi surface remains intact --- both 
the electron and hole Fermi surfaces expands equally. This feature is very much 
in agreement with that of the
work done by Dhaka {\it et al}.,  \cite{Dhakadop}.
 
We provide a microscopic origin of structural transition in BaFe$_{2-x}$Ru$_x$As$_2$. 
Using temperature and doping dependent lattice parameters on Ru doped BaFe$_2$As$_2$
we show through detailed first principle simulations that the electronic structure carries
the `finger prints' of the structural parameters like
 $z_{As}$, Fe-As bond distance and 
reproduces the experimentally observed angle
resolved photo emission spectroscopy data that have so far remained unexplained.
Below structural transition an orbital order develops between d$_{xz}$ and d$_{yz}$ orbitals
of Fe. On the other hand, temperature dependent modifications of d$_{xz}$, d$_{yz}$ bands cause 
loss of nesting causing suppression of spin density wave transition.
Total band energies at high symmetric $\Gamma$ $\&$ X-points of d$_{xz}$, d$_{yz}$ bands
become non-degenerate at structural transition whose temperature dependence is very similar 
to that of the observed two Fe-Fe distances (or $a$ (T) and $b$ (T)); whereas the difference of 
band energies at $\Gamma$ $\&$ X-points of the said bands give rise to orbital order that follows 
the temperature dependence of the orthorhombocity parameter.
Therefore, orbital fluctuations play a dominant role in the magneto-structural transition
in Ru doped BaFe$_2$As$_2$ systems. 
Ru substitution (upto 40 $\%$) do not show 
the nature of charge carrier doping from Fermi surface evolution.
The hole like Fermi surface shrinks with 
temperature but the electron Fermi surface expands comparatively slowly, explains 
the momentum distribution curves observed in ARPES and temperature dependent loss 
of Fermi surface nesting. 
Finally, we demonstrated that the thermal variations of z$_{As}$ obtained from experiments, when used
as inputs in first principle simulation studies, produce realistic theoretical results 
with respect to the electronic structure that is observed experimentally and 
perhaps should be used 
in all families of Fe-based materials in order to  provide better insight.

{\bf Acknowledgements} One of us (SS) acknowledges the HBNI, RRCAT for
financial support and encouragements. We thank Dr. G.
S. Lodha, Dr. P. D. Gupta for their encouragements and support in this work.


\begin{thebibliography}{9}
\bibitem {Kamihara} Y. Kamihara, {\it et al.}, 
 J. Am. Chem. Soc. {\bf 130}, 3296 (2008). 
\bibitem {application} Jun-ichi Shimoyama,  Superconductor Science and Technology, {\bf 27}, 044002 (2014).
\bibitem {stewart} G. R. Stewart, Rev. Mod. Phys.,{\bf 83} 1589 (2011).
\bibitem{Luetkens} H. Luetkens, {\it et al.},
Nat. Mater. {\bf 8}, 305 (2009).
\bibitem {Huang} Q. Huang {\it et al.}, Phys. Rev. Lett. {\bf 101}, 257003 (2008).
\bibitem {Li} S. Li {\it et al.}, Phys. Rev. B
 {\bf 80}, 020504 (2009).
\bibitem {Parkar} D. R. Parkar, {\it et al.},
Chem. Commun. (Cambridge) {\bf 16}, 2189 (2009).
\bibitem{Liu} R. H. Liu {\it et al.}, 
Europhys. Lett. {\bf 94}, 27008 (2011).
\bibitem{Cruz} C. de la Cruz {\it et al.}, Nature {\bf 453} 899 (2008).
\bibitem{Pratt} D. K. Pratt {\it et al.}, Phys. Rev. Lett. {\bf 130} 087001 (2009).
\bibitem{Drew} A. J. Drew {\it et al.}, Nat. Mater. {\bf 8}, 310 (2009).
\bibitem{Mazin} I. I. Mazin {\it et al.}, Phys. Rev. Lett. {\bf 101} 057003 (2008).
\bibitem{Dong} J. Dong {\it et al.}, Europhys. Lett.{\bf 83} 27006 (2008).
\bibitem{orbital} T. Yoshida, S. Ideta {\it et al.}, http://arxiv.org/abs/1301.4818.
\bibitem{nematic} R. M. Fernandes, A. V. Chubukov and J. Schmalian, Nature Phys. {\bf 10} 97 (2014);
E. P. Rosenthal {\it et al.}, Nature Phys. {\bf 10}, 225 (2014).
\bibitem{Mizuguchi} Y. Mizuguchi {\it et al.}, Supercond. Sci. Technol.{\bf 23}, 054013 (2010).
\bibitem{Lee} C. H. Lee {\it et al.}, Solid State Commun.{\bf 152}, 644 (2012).
\bibitem{Zhao} J. Zhao {\it et al.}, Nat Mater.{\bf 12}, 953 (2008).
\bibitem{Shirage} P. M. Shirage {\it et al.}, Physica C {\bf 469}, 355 (2013).
\bibitem{DhakaT} R. S. Dhaka, S. E. Hahn, E. Razzoli, Rui Jiang, M. Shi, B. N. Harmon, A. Thaler, 
S. L. Bud'ko, P. C. Canfield and Adam Kaminsk, Phys. Rev. Lett. {\bf 110},067002 (2013).
\bibitem{DJSingh}L. Zhang and David J. Singh, Phys. Rev. B {\bf 79}, 174530 (2009).
\bibitem{arxiv} S. Sharma {\it et al.}, arXiv 1312.7055 (2013).
\bibitem{dft} Z. P. Yin, S. Lebegue, M. J. Han, B. P. Neal, S. Y. Savrasov, and
W. E. Pickett, Phys. Rev. Lett. {\bf 101}, 047001(2008).
\bibitem{zAs} D. Kasinathan {\it et al.}, New J. Phys. {\bf 11}, 025023 (2009).
\bibitem{lifshitz} I. M. Lifshitz, Zh. Eksp. Teor. Fiz. {\bf 38}, 1569 (1960);
I. M. Lifshitz, Sov. Phys. JETP {\bf 11}, 1130 (1960).
\bibitem{yili} T. Yildirim, Phys. Rev. Lett.{\bf 101}, 057010 (2008).
\bibitem{magno}Kangjun Seo, B. Andrei Bernevig, and Jiangping Hu, Phys. Rev. Lett.
 {\bf 101}, 206404 (2008).
\bibitem{CASTEP} S. J. Clark {\it et al.},
    Zeitschrift fuer Kristallographie {\bf 220}, 567 (2005).
\bibitem{PBE} J. P. Perdew {\it et al.}, Phys. Rev. Lett. {\bf 77}, 3865 (1996).
\bibitem{AFM3} E. Aktrk {\it et al.},
  Phys. Rev. B {\bf 79}, 184523 (2009).
\bibitem{lifshitz1} F. Occelli, D. L. Farber, J. Badro, C. M. Aracne, D. M. Teter, 
M. Hanfland, B. Canny, and B. Couzinet
Phys. Rev. Lett. {\bf 93}, 095502 (2004).
\bibitem{sci-rep} M. Sunagawa {\it et al.}, Nature (scientific report) {\bf 4}, 4381 (2014); 
DOI: 10.1038/srep04381 
\bibitem{ghosh} Haranath Ghosh, H. Purwar, Europhys. Lett. {\bf 98}, 57012 (2012).
\bibitem{Dhakadop} R. S. Dhaka, Chang Liu, R.M. Fernandes, Rui Jiang, C. P. Strehlow, T. Kondo, 
A. Thaler, J\'{o}rg Schmalian, S. L. Budko, P. C. Canfield and A. Kaminski, 
Phys. Rev. Lett. {\bf 107}, 267002 (2011).
\end{thebibliography}
\end{document}